\documentclass[12pt]{article}
\usepackage{amssymb,amsmath}
\usepackage{epsfig} 
\usepackage{times}

\def\S{\hspace{-3pt}\not\!S}

\def\a{\alpha}
\def\b{\beta}
\def\g{\gamma}
\def\d{\delta}

\def\vta{\vartheta}

\def\nn{\nonumber}
\def\be{\begin{equation}}             \def\ee{\end{equation}}
\def\bea{\begin{eqnarray} }           \def\eea{\end{eqnarray}}
             
\def\sS{\hbox{$\not\hspace{-2.5pt}S$}}

\newcommand{\dN}{I\hspace{-3.5pt} H} 
\def\H{\dN}

\begin{document}

\title{On the theory of the skewon field:\\ From electrodynamics to
  gravity}

\author{F.W.\ Hehl,$^{1,2,}$\footnote{email: hehl@thp.uni-koeln.de}
  $\;$ Yu.N.\ Obukhov,$^{1,3,}$\footnote{email: yo@thp.uni-koeln.de}
  $\;$ G.F.\ Rubilar,$^{4,}$\footnote{email: grubilar@udec.cl} $\;$
  M.\ Blagojevi\'c$\,^{5,6,}$\footnote{email: mb@phy.bg.ac.yu}\\ \\ 
  $^1$ Institute for Theoretical Physics, University of Cologne\\ 
  50923 K\"oln, Germany \\ $^2$ Department of Physics and Astronomy,
  University of Missouri-Columbia\\ Columbia, MO 65211, USA\\ $^3$
  Department of Theoretical Physics, Moscow State University\\ 117234
  Moscow, Russia\\ $^4$ Departamento de F{\'{\i}}sica, Universidad de
  Concepci\'on\\ Casilla 160-C, Concepci\'on, Chile\\ $^5$ Institute
  of Physics, P.\ O.\ Box 57, 11001 Belgrade, Serbia\\ $^6$ Department
  of Physics, University of Ljubljana\\ 1000 Ljubljana, Slovenia}

\date{07 June 2005, {\it file skewonfield13.tex}}

\maketitle

\begin{abstract}
  The Maxwell equations expressed in terms of the excitation
  $\H=({\cal H}, {\cal D})$ and the field strength $F=(E,B)$ are
  metric-free and require an additional constitutive law in order to
  represent a complete set of field equations. In vacuum, we call this
  law the ``spacetime relation''. We assume it to be local and linear.
  Then $\H=\H(F)$ encompasses 36 permittivity/permeability functions
  characterizing the electromagnetic properties of the vacuum. These
  36 functions can be grouped into $20+15+1$ functions. Thereof, 20
  functions finally yield the {\it dilaton\/} field and the {\it
    metric\/} of spacetime, 1 function represents the {\it axion\/}
  field, and 15 functions the (traceless) {\it skewon\/} field
  \hbox{$\not\!S_i{}^j$} (S slash), with $i,j=0,1,2,3$.  The
  hypothesis of the existence of \hbox{$\not\!S_i{}^j$} was proposed
  by three of us in 2002. In this paper we discuss some of the
  properties of the skewon field, like its electromagnetic energy
  density, its possible coupling to Einstein-Cartan gravity, and its
  corresponding gravitational energy.
\end{abstract}

\noindent Keywords: Classical electrodynamics, skewon field, 
general relativity, Einstein-Cartan theory, dilaton field, axion
field.


\section{Introduction}

At the centennial of the proposition of special relativity theory by
Einstein (1905), it is worthwhile to remember that Einstein's paper
was ``On the electrodynamics of moving bodies,'' see
\cite{RelativityDover}. The task Einstein had taken up was to develop
a consistent framework for accommodating Maxwell's theory of
electrodynamics as well as classical mechanics. The tool for achieving
this was to study the motion of charged bodies under the action of an
electromagnetic field. Maxwell's theory, suitably interpreted,
survived, Newton's mechanics had to be extended if high relative
velocities were involved.

Thus, Maxwell's theory emerged as a prime example of a special
relativistic field theory that is intrinsically related to the
Poincar\'e group (also known as inhomogeneous Lorentz group).
Accordingly, electrodynamics is conventionally thought to take place
in the flat Minkowski space of special relativity as pointed out by
Minkowski in his geometrical formulation of special relativity in
1908, see \cite{RelativityDover}.

The success of special relativity was so striking that the historical
fact of the close association of electrodynamics with special
relativity stuck in the minds of most physicists and is believed to be
a physical fact --- even though the development of classical
electrodynamics during the last 100 years shows the opposite: The
foundations of electrodynamics have nothing to do with special
relativity and the Poincar\'e group, they are rather of a {\it
  generally covariant\/} (``topological'') nature based on the
conservation laws of electric charge and magnetic flux.

This development started with Einstein \cite{Einstein1916} who shortly
after the publication of his general relativity theory observed
that the Maxwell equations can be formulated in such a way that
neither the metric nor the Christoffel symbols appear in them. In his
notation ($\mu,\nu,...=0,1,2,3$), they read\footnote{Einstein used
  subscripts for denoting the coordinates $x$, i.e., $x_\tau$ etc.
  Moreover, we dropped twice the summation symbols $\Sigma$.}
\begin{equation}\label{akademie}
  \frac{ \partial F_{\rho\sigma}}{\partial x^\tau}+ \frac{ \partial
    F_{\sigma\tau}}{\partial x^\rho}+ \frac{ \partial
    F_{\tau\rho}}{\partial x^\sigma}=0\,,\quad {\cal
    F}^{\mu\nu}=\sqrt{-g}g^{\mu\alpha}
    g^{\nu\beta}F_{\alpha\beta}\,,\quad \frac{\partial{\cal
    F}^{\mu\nu}}{\partial x^\nu}={\cal J}^\mu\,.
\end{equation}
Here we draw on our paper \cite{Okun}.  The Maxwell equations
(\ref{akademie})$_1$ and (\ref{akademie})$_3$ are apparently metric
free.  Moreover, since the excitation ${\cal F}^{\mu\nu} $ is
considered to be a tensor {\it density} of type $2\brack 0 $ and the
field strength $F_{\rho\sigma}$ a tensor of type $0\brack 2 $, both
equations are --- even though only partial derivatives operate in them
--- covariant under general coordinate transformations
(diffeomorphisms). In other words, the system consisting of
(\ref{akademie})$_1$ and (\ref{akademie})$_3$ doesn't couple to the
gravitional potential as long as (\ref{akademie})$_2$ is not
substituted. A similar presentation of Maxwell's equations was given
by Einstein in his ``Meaning of Relativity'' \cite{meaning} in the
part on general relativity.

At first sight, this separation of the Maxwell equations into the
Maxwell equations proper of (\ref{akademie})$_1$ and
(\ref{akademie})$_3$ and the spacetime relation (\ref{akademie})$_2$
may appear to look like a formal trick. However, it is well known that
the electric excitation $\cal D$ and the magnetic excitation $\cal H$
are both directly measurable quantities, see, e.g., Raith
\cite{Raith}. Accordingly, the electromagnetic field is represented
operationally not only by means of the electric and magnetic field
strengths $E$ and $B$ --- both measured via the Lorentz force --- but
also by means of the electric and magnetic excitations $\cal D$ and
$\cal H$.
 
On the foundations of general relativity, besides the equivalence
principle, there lays the principle of general covariance. And the
Maxwell equations (\ref{akademie})$_1$ and (\ref{akademie})$_3$ are
generally covariant {\em and\/} metric independent. Since in general
relativity the metric $g$ is recognized as gravitational potential, it
is quite fitting that the fundamental field equations of
electromagnetism do {\em not\/} contain the gravitational potential.
Consequently, the Maxwell equations in their premetric form are valid
in any 4D differential manifold, provided the latter can be split
locally into 1+3. Accordingly, they are not only beyond special
relativity, but also beyond {\it general\/} relativity.

The point of view that the fundamental structure of electrodynamics
can be best understood because of the existence of conservation
laws that can be formulated generally covariant and metric-free has
been mainly developed by Kottler (1922), \'E.Cartan (1923), and van
Dantzig (1934), see \cite{Birkbook}. Modern presentations of this
``premetric electrodynamics'' have been given, e.g., by
Truesdell-Toupin \cite{TT}, Post \cite{Post}, Kovetz \cite{Attay},
Rubilar \cite{Dr.Guillermo}, Hehl \& Obukhov \cite{Birkbook}, Kiehn
\cite{RMKbook}, Delphenich \cite{Dave1}, and
Lindell\footnote{Lindell's presentation of electrodynamics in the
  framework of exterior differential forms is metric independent.
  However, in order to make himself understood to his engineering
  public, he often interpretes the differential-form expressions in
  terms of metric-dependent Gibbsian vector expressions.} \cite{Ismo},
see also Itin \cite{Yakov1,Yakov2}.


\section{Premetric electrodynamics}

\subsection{The Maxwell equations}

The conservation of electric charge leads to the inhomogeneous Maxwell
equation:
\begin{equation}\label{inhomMax}
  d\H=J \qquad (\partial_j\check{\cal H}^{ij}=\check{J}^i)\,.
\end{equation}
The first version in exterior calculus is written in terms of the
twisted excitation 2-form $\H=\H_{ij}\,dx^i\wedge dx^j/2$ and the
twisted current 3-form $J=J_{ijk}\,dx^i\wedge dx^j\wedge dx^k/6$. The
translation into the corresponding version in components is achieved
by $\check{H}{}^{ij}:=\epsilon^{ijkl}\H_{kl}/2$ and $
\check{J}{}^i:=\epsilon^{ijkl}J_{jkl}/6$, where $ \epsilon^{ijkl}$ is
the totally antisymmetric Levi-Civita symbol with components of value
$\pm 1,0$. Magnetic flux conservation is represented by the
homogeneous Maxwell equation

\begin{center}
\epsfig{file=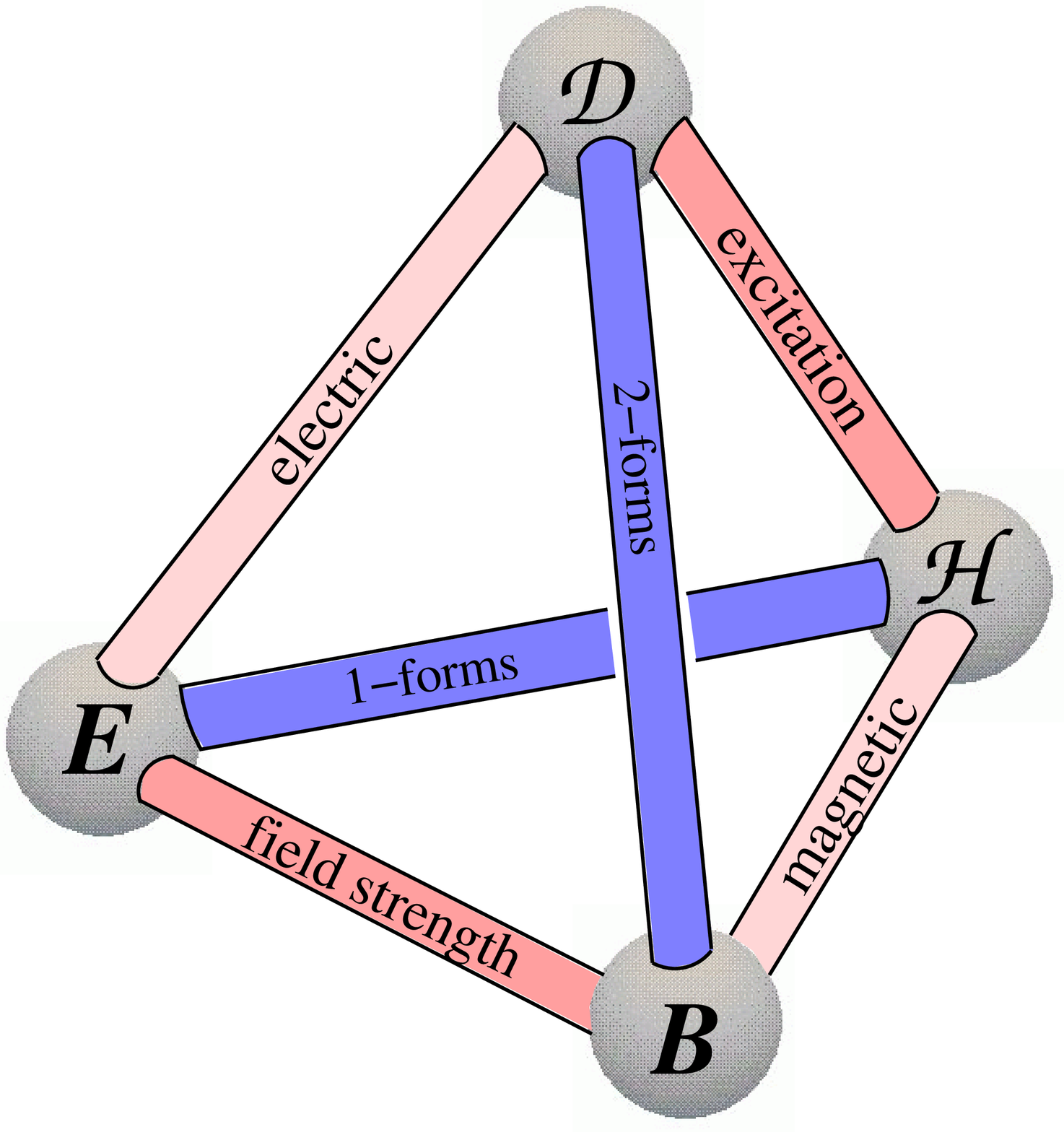,height=7truecm}
\end{center}

Fig.1. {\em The tetrahedron of the electromagnetic field.} The
excitation $\H=({\cal H}, {\cal D})$ and the field strength $F=(E,B)$
are 4-dimensional quantities of spacetime, namely 2-forms with and
without {\it twist}, respectively. They describe the electromagnetic
field completely. Of electric nature are ${\cal D}$ and $E$, of
magnetic nature ${\cal H}$ and $B$. In 3 dimensions, ${\cal H}$ and
$E$ are twisted and untwisted 1-forms, respectively; analogously,
${\cal D}$ and $B$ are twisted and untwisted 2-forms, respectively.
The magnetic and the electric excitations $\H=({\cal H},{\cal D})$ are
extensities, also called quantities (how much?), the electric and the
magnetic field strengths $E$ and $B$ are intensities, also called
forces (how strong?).\bigskip

\begin{equation}\label{homMax}
  dF=0 \qquad (\partial_{[i}F_{jk]}=0)\,,
\end{equation}
with the field strength 2-form $F=F_{ij}\,dx^i\wedge dx^j/2$. The
decompositions into time and space read
\begin{equation}\label{13decomp}
  \H=-{\cal H}\wedge dt+{\cal D}\,,\qquad F=E\wedge dt+B\,,
\end{equation}
compare the scheme in Fig.1. Conservation laws can be reduced to
counting procedures. No distance concept is required in this context,
rather only the ability to circumscribe a definite volume or an area.
As a consequence, no metric occurs anywhere in (\ref{inhomMax}),
(\ref{homMax}), and (\ref{13decomp}).

\subsection{Local and linear spacetime relation}

Excitation and field strength in vacuum (generalization to media is
possible) are assumed to be related by a local and linear relation
\begin{equation}\label{loclin}
  \H=\kappa[F]\quad\qquad(\H_{\a\b}=\frac 12\, \kappa_{\a\b}{}^{\g\d}
  \,F_{\g\d})\,.
\end{equation}
The constitutive tensor $ \kappa_{\a\b}{}^{\g\d}= -\kappa_{\b\a}
{}^{\g\d} =- \kappa_{\a\b}{}^{\d\g}$ has 36 independent components.
These components can be understood in terms of the tensor-valued
2-form
\begin{equation}\label{frakK}
  \mathfrak{K}^{\a\b}=\frac
  12\,\kappa_{\g\d}{}^{\a\b}\,\vta^\g\wedge\vta^\d\,.
\end{equation}
Then (\ref{loclin})$_{1}$ can be written explicitly as
\begin{equation}\label{HfrakK}
  \H=\frac 12\,\mathfrak{K}^{\a\b}e_\b\rfloor e_\a\rfloor F\,.
\end{equation}
The constitutive 2-form $\mathfrak{K}^{\a\b} $ can be split, according
to 36=20+15+1, into three irreducible pieces,
\begin{equation}\label{Kdec}
  \mathfrak{K}^{\a\b}=\underbrace{\,^{(1)}
    \mathfrak{K}^{\a\b}}_{\begin{footnotesize}\hbox{principal}
    \end{footnotesize}}+
  \underbrace{ \,^{(2)} \mathfrak{K}^{\a\b}}_{\begin{footnotesize}
      \hbox{skewon} \end{footnotesize}}+
  \underbrace{\,^{(3)}\mathfrak{K}^{\a\b}}_{\begin{footnotesize}
      \hbox{axion}
    \end{footnotesize}}\qquad ( \kappa_{\a\b}{}^{\g\d}= 
  \sum\limits_{A=1}^{3}\,^{(A)} \kappa_{\a\b}{}^{\g\d})\,.
\end{equation}
A detailed proof is given in the Appendix. In particular,
\begin{equation}
{}^{(2)}\mathfrak{K}^{\alpha\beta} =
\not\!\mathfrak{K}^{[\alpha}\wedge\vartheta^{\beta]},
\qquad {}^{(3)}\mathfrak{K}^{\alpha\beta} = \alpha\,\vartheta^\alpha\wedge
\vartheta^\beta\,.
\end{equation}

If the spacetime relation can be derived from a Lagrangian, then the
skewon piece $^{(2)}\mathfrak{K}^{\alpha\beta}$ has to vanish. On the
other hand, $^{(2)}\mathfrak{K}^{\alpha\beta}$ is a permissible
structure provided it is related to dissipative processes. This is
indeed the case, see \cite{Birkbook}. The hypothesis of the existence
of $^{(2)}\mathfrak{K}^{\alpha\beta}$ was proposed by three of us
\cite{skewonoriginal}. Its effect on the light propagation has been
studied in the meantime \cite{skewon}. The axion piece
$^{(3)}\mathfrak{K}^{\alpha\beta}$ had been proposed much earlier in
an elementary particle context, see the axion electrodynamics of
Wilczek \cite{Wilczek} and the literature given there.

In particular for a comparison with the literature \cite{Post} it is
convenient to introduce the equivalent constitutive tensor density
\begin{equation}\label{chi}
  \chi^{\a\b\g\d}:=\frac 12\,\epsilon^{\a\b\mu\nu}\,
  \kappa_{\mu\nu}{}^{\g\d}\,,\qquad \chi^{\a\b\g\d}=
  \sum\limits_{A=1}^{3}\,^{(A)}\chi^{\a\b\g\d} \,,
\end{equation}
with 36=20+15+1 independent components. Its skewon and its axion
pieces can be mapped to a tensor (15 components) and a pseudo-scalar
(1 component), respectively,
\begin{equation}\label{Sslash}
  \S_\a{}^\b:=\frac 14\,\hat{\epsilon}_{\a\g\d\epsilon}\,^{(2)}
  \chi^{\g\d\epsilon\b}\,,\qquad \a:=\frac {
    1}{4!}\,\hat{\epsilon}_{\a\b\g\d}\,^{(3)} \chi^{\a\b\g\d}\,.
\end{equation}
We have $\S_\a{}^\a=0$. For the 1-form $ \S^\a:=\S_\b{}^\a\vta^\b$,
with $e_\a\rfloor \S^\a=0$, we find by some algebra,
\begin{equation}\label{translate}
  \S^\a=-\frac{1}{ 2}\,\not\!\mathfrak{K}^\a\,,\qquad \a=\frac{1}{12}
  \,\mathfrak{K}\,.
\end{equation}

Up to here, our considerations were premetric. If we put the physical
dimension of $^{(1)} \mathfrak{K}^{\a\b}$ into a function
$\lambda(x)$, the so-called dilaton, then we have 1 dilaton field
$\lambda$, 19 remaining components of the principal part $^{(1)}
\mathfrak{K}^{\a\b}/\lambda$, 15 skewon components
$\hbox{$\S_\a{}^\b$}$, and 1 axion component $\a$. This is as far as
we can go with the premetric concept. The study of the properties of
light propagation can help to constrain the constitutive tensor of
spacetime.

The skewon field, a specific kind of permeability/permittivity of
spacetime, will be in the center of our interest. It is non-Lagrangian
and dissipative, and it diffracts light.  Here we want to address the
problem of how it could couple to gravity. However, first we want to
turn our attention to its electromagnetic energy-momentum.

\section{The electromagnetic energy-momentum density of the 
  skewon field}

The 3-form of the electromagnetic energy-momentum current can be taken
from \cite{Birkbook}, e.g.:
\begin{equation}\label{simax}
 \Sigma_\alpha :={\frac 1 2}\left[F\wedge(e_\alpha\rfloor
    \H) - \H\wedge (e_\alpha\rfloor F)\right]\,.
\end{equation}
We decompose the 3-form according to $\Sigma_\a=\Sigma_{klm\a}
\,dx^k\wedge dx^l\wedge dx^m/6$. Then the corresponding
energy-momentum tensor in tensor calculus can be defined as ${\cal
  T}_i{}^{j}:=\epsilon^{jklm}\,\Sigma_{klm\,i}/6 $ or
\begin{equation}\label{Tij}
  {\cal T}_i{}^{j} =\frac{1}{4}\delta_i^j F_{kl}\check{{\cal
      H}}^{kl}-F_{ik}\check{{\cal H}}^{jk}\,.
\end{equation}

Let us now turn to the skewon part. The excitation can be decomposed
in principal, skewon, and axial parts according to
$\H={}^{(1)}\H+{}^{(2)}\H+{}^{(3)}\H$. Since $\H$ enters (\ref{simax})
linearly, we find $\Sigma_\a= {}^{(1)}\Sigma_\a+{}^{(2)}\Sigma_\a
+{}^{(3)}\Sigma_\a$, with
\begin{eqnarray}
  \Sigma_\alpha|_{\rm skewon}\equiv {}^{(2)}\Sigma_\a& =& {\frac 1
    2}\left[F\wedge(e_\alpha\rfloor {}^{(2)}\H) - {}^{(2)}\H\wedge
    (e_\alpha\rfloor F)\right]\nonumber\\ &=&{\frac 1
    2}e_\alpha\rfloor(F\wedge {}^{(2)}\H) - {}^{(2)}\H\wedge
  e_\alpha\rfloor F\,.\label{whatever}
\end{eqnarray} 
The skewon part of the excitation was derived in (\ref{nun6}) as
\begin{equation}
{}^{(2)}\H=\frac{1}{2}\not\!\!\mathfrak{K}^{\alpha}\wedge
\,e_\alpha\rfloor F\,.
\end{equation} 
We multiply it by $F$, apply the anti-Leibniz rule for the interior
product, and recall that $e_\a\rfloor
\not\!\!\mathfrak{K}^{\alpha}=0$:
\begin{eqnarray}
  F\wedge{}^{(2)}\H&=&\frac{1}{2}\,F\wedge\not\!\!
  \mathfrak{K}^{\alpha}\wedge \,e_\alpha\rfloor F=-\frac
  12\,e_\a\rfloor(F\wedge\not\!\! \mathfrak{K}^\a\wedge F)+\frac
  12\,(e_\a\rfloor F) \wedge\not\!\!\mathfrak{K}^\a\wedge F\nonumber\\ 
  &&+\frac 12\,F\wedge(e_\a\rfloor\not\!\!\mathfrak{K}^\a)\wedge
  F=-\frac 12\,F \wedge\not\!\!\mathfrak{K}^\a\wedge e_\a\rfloor F\,.
\end{eqnarray} 
Thus, $F\wedge {}^{(2)}\H= 0$. We substitute this into
(\ref{whatever}) and find
\begin{equation}\label{skewenergymomentum'}
  { \Sigma_\alpha|_{\rm skewon}=
    (e_\alpha\rfloor F)\wedge
    \S^\beta\wedge(e_\beta\rfloor F)\qquad\hbox{(premetric
    result)}\,.}
\end{equation} 

A similar computation can be performed for the energy-momentum
tensor.  We have (\cite{Birkbook}, p.256)
\begin{equation}\label{crypto2a}
  ^{(2)}\H_{ij}=2\, {\!\not \!S}_{[i}{}^kF_{j]k}\qquad
  \hbox{or}\qquad{}^{(2)}\check{\cal H}^{mn}=\epsilon^{mnij} {\!\not
    \!S}_{i}{}^kF_{jk}\,.
\end{equation} 
On substitution of this into (\ref{Tij}), we find the skewon part of
the energy-momentum tensor as\footnote{If we translate (\ref{skewen1})
  into the energy-momentum 3-form, then, in components, the
  corresponding formula reads:
\begin{equation}\label{emskewon}\nonumber
  \Sigma_{ijk\alpha}|_{\rm skewon}=\frac{3}{4}
  \left(\not\!\kappa_\alpha{}
    ^mF_{m[i}F_{jk]}+\not\!\kappa_{[i}{}^mF_{jk]}F_{\alpha m}+2
    \not\!\kappa_{[i}{}^mF_{j|m}F_{\alpha|k]}\right)\,.
\end{equation}.}
\begin{equation}\label{skewen1}
 {\cal T}_i{}^{j}|_{\rm skewon}=
  \epsilon^{jklm}F_{ik}\, {\!\not \!S}_{l}{}^nF_{nm} \,,
\end{equation}
which is clearly equivalent to (\ref{skewenergymomentum'}) and also of
premetric nature.

\subsection{Trace}

We transvect (\ref{whatever}) with the coframe and recall that
$F\wedge {}^{(2)}\H= 0$:
\begin{equation}\label{trace}
  \vta^\a\wedge {}^{(2)}\Sigma_\alpha=-\vta^\a\wedge {}^{(2)}\H\wedge
  e_\a\rfloor F= -2 F\wedge {}^{(2)}\H=0\,.
\end{equation} 
Thus, the tracelessness is proved:
\begin{equation}\label{trace1}
  \vta^\a\wedge{}^{(2)} \Sigma_\alpha=0\qquad\hbox{(premetric
    result)}\,.
\end{equation}
Equivalently, ${\cal T}_i{}^i=0$. Thus, the skewonic part of the
energy-momentum is tracefree. Since $ \vta^\a\wedge{}
\Sigma_\alpha=0$, we find the analogous property for the principal
part: $ \vta^\a\wedge{} ^ {(1)} \Sigma_\alpha=0 $.

\subsection{Antisymmetric part}

For these considerations we need the existence of a {\it metric}. We
lower the index of the coframe $\vta_\a:=g_{\a\b}\,\vta^\b$, multiply
the energy-momentum from the left, and antisymmetrize:
\begin{equation}\label{antisymm}
  \vta_{[\a}\wedge {}^{(2)} \Sigma_{\b]} =
  -{}^{(2)}\H\wedge\vta_{[\a}\wedge e_{\b]}\rfloor F=\vta_{[\a} \wedge
  (e_{\b]}\rfloor F)\wedge \not\!  {S}^\g\wedge(e_\g\rfloor F)\,.
\end{equation}
This obviously does {\it not\/} vanish. In contrast, in conventional
Maxwell-Lorentz vacuum electrodynamics, we have, of course, $
\vta_{[\a}\wedge \Sigma_{\b]} = 0$, that is, a symmetric
energy-momentum, see \cite{Birkbook}.

Alternatively, one can consider the 2-form $W:=
e^\alpha\rfloor{}^{(2)}\Sigma_\alpha$ which is proportional to the
left hand side of of (\ref{antisymm}), see Itin \cite{YakovGRG}. Then,
\begin{equation}\label{Yakov1}
W=-(e^\alpha\rfloor {}^{(2)}\H)\wedge (e_\alpha\rfloor F)\,.
\end{equation}
Since $\H$ and $F$ are independent, $W$ doesn't vanish in general. For
the Maxwell-Lorentz case, ${}^{(2)}\H$ is zero and, as a consequence,
$W$ vanishes.

Because the electromagnetic skewon energy-momentum has an
antisymmetric piece, it would contribute to the first field equation
of an Einstein-Cartan-Maxwell (with skewon) system. Thus, we have
another (rather indirect) non-Lagrangian type of coupling of the
skewon field to gravity.

\subsection{Energy density}

The energy density becomes (spatial indices are $a,b,c,...=1,2,3$)
\begin{eqnarray}\label{skewen2}
  {\cal T}_0{}^{0}|_{\rm skewon}&=&
  \epsilon^{0klm}F_{0k}\, {\!\not \!S}_{l}{}^nF_{nm}=
  \epsilon^{0abc}F_{0a}\, {\!\not \!S}_{b}{}^nF_{nc}\nonumber \\ &=&
  \epsilon^{0abc}\left(F_{0a}\, {\!\not
      \!S}_{b}{}^0F_{0c}+F_{0a}\, {\!\not \!S}_{b}{}^dF_{dc}
\right)\, .
\end{eqnarray}
The first term in the parenthesis vanishes because of its symmetry in
$a$ and $c$.  Thus,
\begin{eqnarray}\label{skewen3}
  {\cal T}_0{}^{0}|_{\rm skewon}&=&- \epsilon^{0abc}F_{a0}\, {\!\not
    \!S}_{b}{}^dF_{dc}\,.
\end{eqnarray}

Now we can substitute the electric and the magnetic field strengths:
\begin{eqnarray}\label{skewen4}\nonumber
  {\cal T}_0{}^{0}|_{\rm skewon}&=&-E_{a}\, {\!\not
    \!S}_{b}{}^d\epsilon^{abc}\epsilon_{dce}B^e= E_{a}\, {\!\not
    \!S}_{b}{}^a B^b - E_{a}\, {\!\not \!S}_{b}{}^b\,B^a \,.
\end{eqnarray}
We collect the  terms and find 
\begin{equation}\label{skewen6}
  { {\cal T}_0{}^{0}|_{\rm skewon}= \left( {\!\not
        \!S}_{a}{}^b-\delta_a^b\, {\!\not \!S}_{c}{}^c \right) E_b\,{
      B}^a\,.}
\end{equation}
This is an astonishingly simple {\it premetric\/} result. Note that
the second invariant of the electromagnetic field $I_2:=F\wedge
F=-2d\sigma\wedge B\wedge E$ (see \cite{Birkbook}, p.126) enters the
energy expression inter alia.  Recall also that ${\!\not
  \!S}_{c}{}^c=- {\!\not \!S}_{0}{}^0 $.

\subsection{Specialization: The spatially isotropic skewon field}

The spacetime decomposition of the skewon field reads
\begin{equation}\label{prrameterS}
  \!\not\!S_i{}^j= \left(\begin{array}{cc}-s_c{}^c & m^a \\ n_b &
      s_b{}^a \end{array}\right)\,.
\end{equation}
Nieves \& Pal \cite{NP94} chose (in nuclear matter) a spatially
isotropic skewon field according to
\begin{equation}\label{eq30}
  s_a{}^b= \frac{s}{2}\,\delta_a^b\,,\quad m^a=0\,,\quad n_a=0\quad{\rm
    (Nieves\;\&\;Pal)}\,.
\end{equation}
In order to be able to substitute this into (\ref{skewen6}), we compute
\begin{eqnarray}\label{skewen8}
  s_a{}^b-\delta_a^b\,s_c{}^c=\frac{s}{2}\,\delta_a^b-\delta_a^b\,
  \frac{3s}{2} = -s\,\delta_a^b\,.
\end{eqnarray}
We substitute into (\ref{skewen6}) and find
\begin{equation}\label{skewen7}
  { {}^{\rm k}{\cal T}_0{}^{0}|_{\rm skewon\;N\&P}=
  -s\,  E_a\,{ B}^a\,.}
\end{equation}
Hence the energy density here is proportional to the second invariant
$I_2$. Since Nieves \& Pal didn't compute the energy of their skewon
field, we cannot compare (\ref{skewen7}) with earlier results.

A direct check of (\ref{skewen7}) starts from the premetric
electromagnetic energy density
\begin{equation}\label{energ}
  u=\frac 12\left({\cal D}^a\,E_a+{\cal H}_a\,{ B}^a \right)\,.
\end{equation}
For the Nieves \& Pal skewon we have (\cite{Birkbook}, p.262)
\begin{equation}
  {\cal D}^a|_{\rm skewon\;N\&P}= -s\,{B}^a\,,\qquad {\cal
    H}_a|_{\rm skewon\;N\&P}= -s\,E_a\,.\label{DH-new1a}
\end{equation}
Thus,
\begin{equation}\label{energ1}
  u|_{\rm skewon\;N\&P}= -s\,E_a\,{ B}^a \,,\quad\hbox{q.e.d.}
\end{equation}

\section{Einstein-Cartan theory with skewon, dilaton, and axion interaction}

In electrodynamics, one can think of the constitutive 2-form
$\mathfrak{K}^{\alpha\beta}$ either as a field determined by the
electromagnetic properties of some fixed distribution of background
matter or as a property of spacetime itself.  Whereas the standard
matter fields and the electromagnetic potential are dynamical
variables, $\mathfrak{K}$ is a fixed, non-dynamical (or external)
field.

One can try to describe this situation by an ``effective" Lagrangian
formalism.  Consider, for instance, the simple Lagrangian, quadratic
in the field strengths, ${\cal L}'\sim \chi^{ijkl}F_{ij}F_{kl}$. It is
clear that here only the piece of $\chi^{ijkl}$ symmetric under the
exchange $(i,j)\leftrightarrow(k,l)$ survives and thus the related
field equations do not contain the skewon piece of $\chi$. In a way,
this result might have been expected: the skewon field is known to
cause dissipative effects in electrodynamics \cite{skewon} and,
consequently, one does not expect to have a simple local Lagrangian
description of the complete dynamics.

In trying to extend our understanding of $\mathfrak{K}$ to the
gravitational sector, we adopt the interpretation of $\mathfrak{K}$ as
a {\it property of spacetime\/}, and we will study some of its
consequences.

\subsection{Specialization: principal part with metric and dilaton}

Although it is known that one can construct a gravitational theory
without a metric, all such models are limited to the vacuum case, see,
e.g., \cite{pureconn1,pureconn2,pureconn3,pureconn4}. It is unclear
whether one can construct a viable gravity theory without a metric in
the presence of nontrivial matter sources.  Accordingly, we will now
specialize to the case when the metric field is available as, e.g., in
metric-affine gravity (MAG), see \cite{PRs}. Then, the Maxwell-Lorentz
electrodynamics yields the principal part of the form:
\begin{equation}
  {}^{(1)}\mathfrak{K}^{\alpha\beta} =
  \lambda\,\eta^{\alpha\beta}\qquad
  ({}^{(1)}\kappa_{\g\d}{}^{\a\b}=\lambda\,
  \hat{\epsilon}_{\g\d\mu\nu}\,\sqrt{-g}\,
  g^{\a\mu}g^{\b\nu})\,.\label{1part}
\end{equation}
Here $\eta^{\alpha\beta} =
{}^\star(\vartheta^\alpha\wedge\vartheta^\beta)$ is defined with the
help of the Hodge star $^\star$ for the spacetime metric $g$, whereas
the scalar field $\lambda(x)$ is the dilaton field that represents a
factor of the principal part of $\mathfrak{K}$ absorbing its physical
dimension. The dilaton comes as a companion of the skewon and the
axion even on the premetric level. When $\lambda =$ const, and the
skewon and axion are absent, we recover from (\ref{1part}) the
standard Maxwell-Lorentz electrodynamics.

Now, we recall that the Einstein-Cartan theory (see Blagojevi\'c
\cite{Milutin}, Gronwald et al.\ \cite{gron96}, and/or Trautman
\cite{Trautman}) is determined by the Lagrangian 4-form (in units
$\kappa=8\pi G=1$)
\begin{equation}
V_{\rm EC} = {\frac 12}\,\eta^{\alpha\beta}\wedge R_{\alpha\beta}.
\end{equation}
Noticing that the constitutive 2-form (\ref{frakK}) {\it with the
  principal part\/} (\ref{1part}) provides a natural extension of
$\eta^{\alpha\beta}$ by taking into account the electromagnetic
companions of the metric, we can propose a generalization of the
Einstein-Cartan theory by means of the Lagrangian
\begin{eqnarray}
  V_{\rm gEC}& =& {\frac 12}\,\mathfrak{K}^{\alpha\beta}\wedge
  R_{\alpha\beta}= {\frac 12}\,^{(1)}\mathfrak{K}^{\alpha\beta}\wedge
  {}^{(6)}R_{\a\b} \nonumber\\ && +\,{\frac
    12}\,^{(2)}\mathfrak{K}^{\alpha\beta}\wedge({}^{(2)}R_{\a\b}+
  {}^{(5)}R_{\a\b}) + {\frac 12}\,^{(3)}\mathfrak{K}^{\alpha\beta}
  \wedge\,^{(3)}R_{\a\b}\,.\label{V}
\end{eqnarray}
Here we substituted the irreducible decomposition of the curvature
into 6 pieces $R_{\a\b}=\sum_{A=1}^6{}^{(A)}R_{\a\b}$, see
\cite{PRs}. Since $^{(2)}R_{\a\b}$ is the so-called paircommutator and
$^{(5)}R_{\a\b}$ corresponds to the antisymmetric piece of the Ricci
tensor, we recognize that in (\ref{V}) the skewonic part
$^{(2)}\mathfrak{K}^{\alpha\beta}$ couples only to these specific
post-Riemannian pieces of the curvature. More generally, the
contributions of the skewon and axion are only nontrivial for a
Riemann-Cartan geometry with a nonvanishing torsion 2-form
$T^\alpha$. 

The torsion itself can be also irreducibly decomposed according to
$T^\a={}^{(1)}T^\a+{}^{(2)}T^\a+{}^{(3)}T^\a$, with the second and the
third irreducible torsion pieces defined as usual by
\begin{equation}
{}^{(2)}T^\alpha = {\frac 13}\,\vartheta^\alpha\wedge T,\qquad
{}^{(3)}T^\alpha = -\,{\frac 13}\,{}^\star(\vartheta^\alpha\wedge P).
\end{equation}
The 1-forms of the trace and the axial trace of torsion are introduced
by $T := e_\alpha\rfloor T^\alpha$ and $P :=
{}^\star(\vartheta^\alpha\wedge T_\alpha)$, respectively. By making
use of the first Bianchi identity $DT^\alpha = R_\beta{}^\alpha
\wedge\vartheta^\beta$, we can rewrite the above Lagrangian (\ref{V})
into an equivalent form
\begin{equation}
  V_{\rm gEC} = {\frac 12}\,\lambda\,\eta^{\alpha\beta}\wedge {}^{(6)}
  R_{\alpha\beta} + (D{\S}^\alpha)\wedge({}^{(1)}
  T_\alpha+{}^{(2)}T_\a) - \frac{1}{2}D(\alpha\vartheta^\alpha)\wedge
  {}^{(3)}T_\alpha + d\,\Psi.
\end{equation}
Here the total derivative term contains the 3-form $\Psi := -\,{\!\not
  \!S}^\alpha\wedge ({}^{(1)} T_\alpha+{}^{(2)}T_\a) + {\frac
  12}\alpha\, \vartheta^\alpha\wedge {}^{(3)}T_\alpha$. Obviously, the
skewon field $\S^\a$ couples to the tensor and the vector pieces of
the torsion, the axion field $\a$, however, to the axial torsion
(totally antisymmetric torsion).

\subsection{Generalized gravitational field equations}

The general framework for the derivation of the field equations is
provided by the Noether-Lagrange machinery developed in the review
paper \cite{PRs}, see its Sec.\ 5.8.1. The gravitational field
equations are given by the system of the so-called first and the
second field equations of gravity:
\begin{eqnarray}
DH_\alpha - E_\alpha &=& \Sigma_\alpha,\\
DH_{\alpha\beta} + \vartheta_{[\alpha}\wedge H_{\beta]} &=&
\tau_{\alpha\beta}.
\end{eqnarray}
The sources arise as the variational derivatives of the material
Lagrangian with respect to the coframe and the connection, and they
represent the canonical energy-momentum current $\Sigma_\alpha$ and
the spin current $\tau_{\alpha \beta}$, respectively.

For the Lagrangian (\ref{V}) of the generalized Einstein-Cartan theory
we find straightforwardly the gravitational gauge field momenta
\begin{equation}
  H_\alpha = -\,{\frac {\partial V_{\rm gEC}}{\partial T^\alpha}} =
  0,\qquad H_{\alpha\beta} = -\,{\frac
    {\partial V_{\rm gEC}}{\partial R^{\alpha\beta}}} = -\,{\frac
    12}\,\mathfrak{K}_{\alpha\beta},
\end{equation}
and the gravitational canonical energy 3-form
\begin{equation}
E_\alpha = {\frac 12}\,(e_\alpha\rfloor\mathfrak{K}_{\beta\gamma})
\wedge R^{\beta\gamma}.
\end{equation}
Accordingly, the gravitational field equations read
\begin{eqnarray}
-\,{\frac 12}\,(e_\alpha\rfloor\mathfrak{K}_{\beta\gamma})\wedge 
R^{\beta\gamma} &=& \Sigma_\alpha,\\
-\,{\frac 12}\,D\mathfrak{K}_{\alpha\beta} &=& \tau_{\alpha\beta}.
\end{eqnarray}
Explicitly, the {\it first\/} equation has the form
\begin{equation}
-\,{\frac 12}\,\lambda\,\eta_{\alpha\beta\gamma}\wedge R^{\beta\gamma}
- {\!\not \!S}^\beta\wedge R_{\beta\alpha} + (e_\alpha \rfloor{
\!\not\!S}^\beta)\,\vartheta^\gamma\wedge R_{\beta\gamma} -
  \alpha\,\vartheta^\beta\wedge R_{\alpha\beta} = \Sigma_\alpha.
\end{equation}
Besides the first Einsteinian term (modified by the scalar dilaton
coupling \`a la Brans-Dicke), we now see that the skewon and the axion
fields bring into the first equation new terms which all depend on the
Riemann-Cartan curvature.

The {\it second\/} field equation determines the spacetime torsion in
terms of the spin current of matter {\it and} the additional
contributions of skewon, axion, and dilaton. Explicitly, we have:
\begin{equation}\label{2eq}
-\,{\frac \lambda 2}\,\eta_{\alpha\beta\gamma}\wedge T^\gamma - {\frac 12}
\,\eta_{\alpha\beta}\wedge d\lambda + T_{[\alpha}\wedge\!\not\!S_{\beta]}
- \vartheta_{[\alpha}\wedge D\!\not\!S_{\beta]} - {\frac 12}\,\vartheta_\alpha
\wedge\vartheta_\beta\wedge d\alpha - 
\alpha\,T_{[\alpha}\wedge\vartheta_{\beta]} = \tau_{\alpha\beta}.
\end{equation}
In principle, one can resolve this algebraic equation with respect to the
components of the torsion. In vacuum, when the matter spin vanishes, we
find that the torsion is determined exclusively by the metric companion
fields: the dilaton, the skewon, and the axion (and their derivatives).

There is a particular exact solution of the field equations in vacuum,
which corresponds to a {\it teleparallel geometry\/}. Indeed, for
$\Sigma=\tau=0$, we see that $R_{\alpha\beta} =0$ solves the first
field equation, while (\ref{2eq}) defines the intrinsic torsion of
spacetime in terms of its $\mathfrak{K}$-structure (dilaton, skewon,
and axion).

\subsection{Simple vacuum solution}

Unfortunately, although the second field equation looks rather simple,
it is not easy to find the torsion components from it explicitly.
Nevertheless, we can illustrate how the theory works in a simple case
when the skewon is absent. Then, the terms with $\hbox{$\not\!S$}$
disappear from the equation, and we find that the {\it vacuum} torsion
has the following simple form,
\begin{equation}
T^\alpha = {}^{(2)}T^\alpha + {}^{(3)}T^\alpha\,,
\end{equation}
that is, the tensor piece $^{(1)}T^\a$ of the torsion vanishes.

{}From (\ref{2eq}), we find:
\begin{eqnarray}
T &=& {\frac {3/2}{\lambda^2 + \alpha^2}}\,(\lambda\,d\lambda +
\alpha\,d\alpha),\\
P &=&{\frac {3}{\lambda^2 + \alpha^2}}\,(\alpha\,d\lambda -
\lambda\,d\alpha).
\end{eqnarray}
This can be verified if we notice that the following identities hold
true in exterior calculus:
\begin{eqnarray}
\eta_{\alpha\beta\gamma}\wedge {}^{(2)}T^\gamma = -\,{\frac 23}
\,\eta_{\alpha\beta}\wedge T, &\quad&
\eta_{\alpha\beta\gamma}\wedge {}^{(2)}T^\gamma = {\frac 13}
\,\vartheta_\alpha\wedge\vartheta_\beta\wedge P,\\
{}^{(2)}T_{[\alpha}\wedge\vartheta_{\beta]} = -\,{\frac 13}
\,\vartheta_\alpha\wedge\vartheta_\beta\wedge T, &\quad&
{}^{(3)}T_{[\alpha}\wedge\vartheta_{\beta]} = -\,{\frac 16}
\,\eta_{\alpha\beta}\wedge P.
\end{eqnarray}
We thus conclude that the trace and the axial trace 1-forms of the
torsion are determined, in vacuum, by the dilaton and the axion
fields. In particular, when the axion is absent, $\alpha = 0$, we
recover a Brans-Dicke type gravity constructed in the Einstein-Cartan
framework in \cite{dil1,dil2,dil3}. In that case the axial trace
vanishes, whereas the torsion trace is proportional to the gradient of
the dilaton field. Otherwise, for the case of a constant dilaton,
$\lambda = \lambda_0 =$const, both torsion 1-forms are nontrivial and
depend on the axion only.

Now, when we return to the general case, it is straightforward to verify
that the nontrivial skewon induces the trace-free irreducible part of
torsion, ${}^{(1)}T^\alpha$, in addition to the trace and the axial trace
1-forms.

\section{Gravitational energy in the generalized Einstein-Cartan theory}

We now wish to study the energy of the generalized Einstein-Cartan
theory (\ref{V}) in the framework of the canonical formalism
\cite{Milutin}. Using simple algebra, we can rewrite the gravitational
piece of (\ref{V}) in the form $V_{\rm gEC} = {\cal L}_{\rm
  gEC}\,d^4x$, with \be {\cal L}_{\rm gEC} =
\frac{1}{4}{\chi}_{\a\b}{}^{ij} R_{ij}{}^{\a\b}(\Gamma) \,.  \ee Here,
\bea &&\chi_{\a\b}{}^{ij}= 2\lambda\,\sqrt{-g}\,e^i{}_{[\a}e^j{}_{\b]}
+\bar\chi_{\a\b}{}^{ij}\, , \\ 
&&\bar\chi_{\a\b}{}^{ij}=e_{k\a}e_{l\b}\left(2\epsilon^{klm[i}\sS_m{}^{j]}
  +\a\,\epsilon^{klij} \right)\,.  \eea The Latin and Greek indices
are raised and lowered with the help of the spacetime metric
$g_{ij}=e_i{}^\a e_j{}^\b\,g_{\a\b}$ and the Lorentz metric $g_{\a\b}=
{\rm diag}(+1,-1,-1, -1)$, respectively.  Primary constraints are
similar to those of the standard Einstein-Cartan theory: \bea
&&\pi_\a{^0}\approx 0\, ,\qquad \pi_{\a\b}{^0}\approx 0\, ,\nn\\ 
&&\pi_\a{^a}\approx 0\, ,\qquad
\phi_{\a\b}{^a}:=\pi_{\a\b}{^a}-{\chi}_{\a\b}{}^{0a}\approx 0\,.\nn
\eea Since the Lagrangian is linear in the velocities $\dot\Gamma$, the
canonical Hamiltonian is given by $\mathfrak{H}_c=-{\cal L}_{\rm
  gEC}(\dot\Gamma=0)$: \be \mathfrak{H}_c=
-\frac{1}{4}{\chi}_{\a\b}{}^{ab}R_{ab}{}^{\a\b}
+\frac{1}{2}\Gamma_0{}^{\a\b}\nabla_a{\chi}_{\a\b}{}^{0a} +\partial_a
U^a \, , \ee where $U^a={\chi}_{\a\b}{}^{0a}\Gamma_0{}^{\a\b}/2$.
Looking at the form of $\chi_{\a\b}{}^{ij}$, we see that the canonical
Hamiltonian contains not only the standard Einstein-Cartan piece,
modified by the presence of dilaton, but also an additional
skewon-axion contribution, \be \mathfrak{H}^{\rm SA}_c=
-\frac{1}{4}{\bar\chi}_{\a\b}{}^{ab}R_{ab}{}^{\a\b}
+\frac{1}{2}\Gamma_0{}^{\a\b}\nabla_a{
  \bar\chi}_{\a\b}{}^{a0}+\partial_aU^a_{\rm SA}\, , \ee with
$U^a_{\rm SA}={\bar\chi}_{\a\b}{}^{0a}\Gamma_0{}^{\a\b}/2$.  The total
Hamiltonian has the form \be
\mathfrak{H}_T=\mathfrak{H}_c+u^\a{_0}\,\pi_\a{^0}
+\frac{1}{2}u^{\a\b}{_0}\,\pi_{\a\b}{^0} +u^\alpha{_a}\,\pi_\a{^a}
+\frac{1}{2}u^{\a\b}{_a}\,\phi_{\a\b}{^a}\,.  \ee The simple Hamiltonian
structure obtained so far is sufficient to derive the canonical
expression for the gravitational energy.

In the Hamiltonian formalism, symmetry properties of a dynamical
system are described by the canonical generators $G[\varphi,\pi]$.
Since they act on basic dynamical variables via Poisson brackets,
they must be differentiable. A local functional $F[\varphi,\pi]= \int
d^3x f(\varphi,\partial\varphi,\pi,\partial\pi)$ has well defined
functional derivatives if its variation can be written in the form
$\delta F[\varphi,\pi]=\int d^3 x(A\delta\varphi+B\delta\pi)$, where
terms of the form $\delta(\partial\varphi)$ and $\delta(\partial\pi)$
are absent. If the generator $G[\varphi,\pi]$ is not differentiable,
its form can be improved by adding a suitable surface term, whereupon
it becomes differentiable. On shell, these surface terms represent
the values of the related conserved charges.

The canonical generator of time translations is defined by the total
Hamiltonian:
\be
P_0=\int d^3 x\,\hat{\mathfrak{H}}_T\, ,\qquad
\mathfrak{H}_T\equiv\hat{\mathfrak{H}}_T+\partial_\alpha U^\alpha\, .
\ee
Looking at the skewon-axion piece of $P_0$, we find that its
variation has the form
$$
\delta P^{\rm SA}_0=\int d^3 x\, \delta\hat{\mathfrak{H}}^{\rm SA}_c+N
  =-\frac{1}{2}\int d^3 x\partial_a \left(
   \bar\chi_{\a\b}{}^{ab}\delta\Gamma_b{}^{\a\b}\right)+N\, ,
$$
where $N$ denotes well defined, normal (regular) terms. Thus, $P^{\rm SA}_0$ 
can be made differentiable by adding a surface term:
\bea
&&P_0^{\rm SA}\to\tilde P_0^{\rm SA}=P_0^{\rm SA}+{\cal E}^{\rm SA}\, , 
              \nn\\
&&{\cal E}^{\rm SA}=\frac{1}{2}\int dS_a\left(
     \bar\chi_{\a\b}{}^{ab}\Gamma_b{}^{\a\b}\right)\, .
\eea

In order to ensure the convergence of the surface integral ${\cal
  E}^{\rm SA}$, we have to adopt suitable asymptotic conditions. For
localized gravitational sources (matter fields decrease sufficiently
fast at large distances and give no contribution to surface
integrals), we can assume that spacetime is {\it asymptotically
  flat\/}. The related asymptotic conditions for the gravitational
variables, when expressed in the standard spherical coordinates, take
the simple form: \be e_i{}^\alpha=\delta_i^\a+{\cal O}(1/r)\, ,\qquad
\Gamma_i{}^{\a\b}={\cal O}(1/r^2)\, .  \ee Hence, ${\cal E}^{\rm SA}$
is convergent if \be \bar\chi_{\a\b}{}^{ab}\to{\rm const.}\quad {\rm
  for}\quad r\to\infty\, .  \ee The surface term ${\cal E}^{\rm SA}$
represents the value of the {\it skewon-axion contribution to the
  gravitational energy\/}. It is produced by the interaction between
the skewon-axion term $\bar\chi$, and the connection $\Gamma$. One
should remember that the complete gravitational energy contains also
the standard Einstein-Cartan piece, modified by the presence of the
dilaton. The adopted asymptotics, extended naturally to the dilaton
field by
$$
\lambda(x)\to {\rm const.}\quad {\rm for}\quad r\to\infty\, ,
$$
ensures the conservation of the gravitational energy.

\section{Concluding remarks}
 
(1) According to its definition, the skewon field
is some kind of permeability/permittivity of spacetime --- and this in
a premetric setting when the metric has not yet "condensed". In this
sense, the skewon field is an elementary electromagnetic property of
spacetime.  As such, it influences light propagation.  \medskip

\noindent (2) The skewon field contributes non-trivially to the
electromagnetic energy. In particular, it induces an {\it
  asymmetric\/} electromagnetic energy-momentum tensor, which can
cause specific gravitational effects as a source term in the
Einstein-Cartan-Maxwell system (with skewon).  \medskip

\noindent (3) A smooth deformation of the Einstein-Cartan theory has been
introduced and studied as a simple dynamical model incorporating
gravitational effects of the skewon field. We found the generalized
gravitational field equations and were able to determine the
contribution of the skewon field to the gravitational energy.

\subsection*{Acknowledgments} 
We are very grateful to Yakov Itin (Jerusalem) and to Christian
Heinicke (Cologne) for various helpful remarks.  G.\ Rubilar would
like to thank the Fundacion Andes (Convenio C-13860) for financial
support.

\section*{Appendix. Decomposition of the local and linear 
  constitutive law}

We start with a local (in space and in time) and linear constitutive
law
\begin{equation}\label{operator1}
  \H=\kappa[F]\,.
\end{equation}
The operator $\kappa$ acts on the electromagnetic field strength $F$.
Because of its linearity, we have for any 2-forms $\Psi,\Phi$, if
$a,b$ are two real scalar factors,
\begin{equation}\label{operator1a}
  \kappa[a \Psi+b \Phi]=a\,\kappa[\Psi]+b\,\kappa[\Phi]\,.
\end{equation}
We substitute the decomposition of the field
strength $F$ into (\ref{operator1}):
\begin{equation}\label{operator2}
  \H=\kappa[\frac{1}{2}F_{\a\b}\vta^\a\wedge \vta^\b]=\frac
  12\,\kappa[\vta^\a\wedge \vta^\b]\,F_{\a\b}\,.
\end{equation}
We introduce the constitutive tensor-valued 2-form
\begin{equation}\label{operator2a}
\frak{K}^{\a\b}:=\kappa[\vta^\a\wedge\vta^\b]
\end{equation}
and recall the decomposition $F_{\a\b}=e_\b\rfloor e_\a\rfloor F$.
Then the constitutive relation can be brought into the compact form
\begin{equation}\label{operator3}
  \H=\frac{1}{2}\,\frak{K}^{\a\b}e_\b\rfloor e_\a\rfloor
    F\,,\qquad\hbox{with}\qquad\mathfrak{K}^{\a\b}=-\mathfrak{K}^{\b\a}\,.
\end{equation}
Here the 2-form $\mathfrak{K}^{\a\b}$ decomposes as
\begin{equation}\label{nun1a}
\mathfrak{K}^{\a\b}=\frac{1}{2}\,\kappa_{\g\d}{}^{\a\b}\,
  \vartheta^\g\wedge \vartheta^\d\,.
\end{equation}
Contractions yield a 1-form and a 0-form, respectively:
\begin{equation}\label{nun2}
  \mathfrak{K}^\beta := e_\alpha\rfloor\mathfrak{K}^{\alpha\beta}
  =\kappa_{\alpha\delta}{}^{\alpha\beta}\vartheta^\delta
  =:\kappa_\d{}^\b\,\vartheta^\delta\,,\qquad
  \mathfrak{K}:=e_\b\rfloor\mathfrak{K}^{\b}=\kappa_\b{}^\b=:\kappa\,.
\end{equation}
The tracefree part of the 1-form is
\begin{equation}\label{nun3}
  \not\!\!\mathfrak{K}^{\a}:=\mathfrak{K}^{\a}-\frac
  14\,\mathfrak{K}\vartheta^\a\,.
\end{equation}
In this way we can decompose the constitutive antisymmetric tensor
valued 2-form into its 3 irreducible pieces,
\begin{equation}\label{nun4}
 \mathfrak{K}^{\a\b}=\,^{(1)}\mathfrak{K}^{\a\b}+
  \,^{(2)}\mathfrak{K}^{\a\b}+\,^{(3)}\mathfrak{K}^{\a\b}\,,
\end{equation}
with the skewon and the axion pieces
\begin{equation}\label{nun4a}
  ^{(2)}\mathfrak{K}^{\a\b}:=
  \not\!\!\mathfrak{K}^{[\a}\wedge\vartheta^{\b]}\qquad\hbox{and}\qquad
  ^{(3)}\mathfrak{K}^{\a\b}:=\frac{1}{ 12}\,\mathfrak{K}\,
  \vartheta^\a\wedge\vartheta^\b\,.
\end{equation}
The factors can be determined with some trivial algebra. Note the
constraints
\begin{equation}\label{nun5}
  e_\a\rfloor \,^{(1)}\mathfrak{K}^{\a\b}=0\qquad{\rm and}\qquad
 e_\a\rfloor \,  \not\!\!\mathfrak{K}^{\a}=0\,.
\end{equation}
The irreducible pieces in (\ref{nun4a}) can also be written in
components. With the help of the generalized Kronecker delta (see
\cite{Birkbook}), we find
\begin{equation}\label{nun5x}
  ^{(2)}\kappa_{\g\d}{}^{\a\b}=
  2\delta_{\g\d}^{\varepsilon[\a}\,\S_\varepsilon{}^{\b]} \,,\qquad
  ^{(3)}\kappa_{\g\d}{}^{\a\b}=\delta_{\g\d}^{\a\b}\,\a\,.
\end{equation}

If we substitute (\ref{nun4}) into (\ref{operator3}) and observe
$\vartheta^\a\wedge\left(e_\a\rfloor \omega \right)= p\omega$, where
$\omega$ is a $p$-form, then we finally have
\begin{eqnarray}\label{nun6}
  \H&=& {}^{(1)}\H+ {}^{(2)}\H+ {}^{(3)}\H\nonumber\\ &=&\frac
    12\left(\,^{(1)}\mathfrak{K}^{\a\b}\,e_\b\rfloor e_\a\rfloor +
      \not\!\!\mathfrak{K}^{\a}\wedge \,e_\a\rfloor +\frac 16\,
      \mathfrak{K} \right)F\,.
\end{eqnarray}
Thus, the {\it principal\/} part of the constitutive 2-form
$\mathfrak{K}^{\a\b} $ is represented by the $[_0^2]$ antisymmetric
tensor-valued 2-form $^{(1)}\mathfrak{K}^{\a\b}=
\,-^{(1)}\mathfrak{K}^{\b\a}$, the {\it skewon\/} part by the
vector-valued 1-form $ \not\!\!\mathfrak{K}^{\a}$, and the {\it
  axion\/} part by the pseudoscalar $\mathfrak{K}$. The translation
into our usual language is made by $ \not\!{S}^{\a}= -\frac 12
\not\!\!\mathfrak{K}^{\a}$ and
$\alpha=\frac{1}{12}\,\mathfrak{K}\,.$ 
\medskip

Incidentally, the {IB-medium} of Lindell \cite{IsmoIB} is defined by
$\,^{(1)}\mathfrak{K}^{\a\b}=0$. If additionally
$\,^{(2)}\mathfrak{K}^{\a\b}=0$ (vanishing skewon field), then only
$\,^{(3)}\mathfrak{K}^{\a\b}=\frac{1}{12}\, \mathfrak{K}\,
\vartheta^\a\wedge\vartheta^\b$ is left over, the axion field with 1
component, or, in the language of Lindell \& Sihvola
\cite{LindSihv2004a}, the perfect electromagnetic conductor (PEMC).

\begin{footnotesize}

\centerline{=========}
\end{footnotesize}


\begin{thebibliography}{99}

\bibitem{RelativityDover} H.A.~Lorentz, A.~Einstein, H.~Minkowski and
  H.~Weyl, {\it The Principle of Relativity}, a collection of original
  memoirs on the special and general theory of Relativity, translated
  from the German (Dover: New York, 1952).

\bibitem{Einstein1916} A.\ Einstein, {\it Eine neue formale Deutung
    der Maxwellschen Feld\-glei\-chungen der Elektrodynamik} (A new
    formal interpretation of Maxwell's field equations of
    electrodynamics), {\sl Sitzungsber.\ K\"onigl.\ Preuss.\ Akad.\
    Wiss.\ Berlin} (1916) pp.  184--188; see also {\em The collected
    papers of Albert Einstein.} Vol.6, A.J.\ Kox et al., eds. (1996)
    pp. 263--269.

  \bibitem{Okun} F.W.~Hehl and Yu.N.~Obukhov, {\it Dimensions and
      units in electrodynamics,} Gen.\ Rel.\ Grav. {\bf 37} (2005)
    733--749; arXiv.org/physics/0407022.

\bibitem{meaning} A.\ Einstein, {\it The Meaning of Relativity,} 5th
  ed.\ (Princeton University Press: Princeton, 1955).

\bibitem{Raith} W.\ Raith, ed., {\em Bergmann-Schaefer, Lehrbuch der
    Experimentalphysik, Vol.2, Elektromagnetismus}, new 9th ed. (de
    Gruyter: Berlin, to be published in autumn 2006).

\bibitem{Birkbook} F.W.~Hehl and Yu.N.~Obukhov, {\em Foundations of
      Classical Electrodynamics --- Charge, Flux, and Metric}.
    Birk\-h{\"a}user, Boston (2003).

\bibitem{TT} C. Truesdell and R.A. Toupin: {\it The Classical Field
    Theories}, in: {\it Handbuch der Physik}, S. Fl\"ugge ed., Vol.
  III/1 (Springer, Berlin 1960) pp.\ 226--793.

\bibitem{Post} E.J.\ Post, {\it Formal Structure of Electromagnetics
    -- General Covariance and Electromagnetics} (North Holland:
  Amsterdam, 1962, and Dover: Mineola, New York, 1997).

\bibitem{Attay} A.\ Kovetz, {\it Electromagnetic Theory} (Oxford
  University Press: Oxford, 2000).

\bibitem{Dr.Guillermo} G.F.\ Rubilar, {\it Linear pre-metric
    electrodynamics and deduction of the lightcone\/}, {\sl Ph.D.
    Thesis} (University of Cologne, June 2002); see {\sl Ann.\ Phys.\ 
    (Leipzig)} {\bf 11} (2002) 717--782.

\bibitem{RMKbook} R.M. Kiehn, {\it Non-equilibrium and irreversible
    electrodynamics,} November 2003 (76 pages), see

  {\tt http://www22.pair.com/csdc/car/carhomep.htm} .

\bibitem{Dave1} D.H.~Delphenich, {\it On the axioms of topological
    electromagnetism,} Ann.\ Phys.\ (Leipzig) {\bf 14} (2005)
  347--377; updated version of arXiv.org/hep-th/0311256.

\bibitem{Ismo} I.V.~Lindell, {\it Differential Forms in
    Electromagnetics.} IEEE Press, Piscataway, NJ, and
  Wiley-Interscience (2004).

\bibitem{Yakov1} Y.~Itin, {\it Caroll-Field-Jackiw electrodynamics in
    the pre-metric framework,} Phys.\ Rev. {\bf D70} (2004) 025012 (6
  pages); arXiv.org/hep-th/0403023.

\bibitem{Yakov2} Y.~Itin and F.~W.~Hehl, {\it Is the Lorentz signature
    of the metric of spacetime electromagnetic in origin?,} Annals
  Phys.(NY) {\bf 312} (2004) 60-83; arXiv.org/gr-qc/0401016.

\bibitem{skewonoriginal} F.W.\ Hehl, Yu.N.\ Obukhov, G.F.\ Rubilar,
  {\em On a possible new type of a T odd skewon field linked to
    electromagnetism}. In: {\sl Developments in Mathematical and
    Experimental Physics,} A.\ Macias, F.\ Uribe, and E.\ Diaz, eds.
  Volume A: Cosmology and Gravitation (Kluwer Academic/Plenum
  Publishers: New York, 2002) pp.241-256; arXiv.org/gr-qc/0203096.

\bibitem{skewon} Yu.N.~Obukhov and F.W.~Hehl, {\it Possible skewon
    effects on light propagation}, Phys.\ Rev. {\bf D70} (2004) 125015
  (14 pages); arXiv.org/physics/0409155.

\bibitem{Wilczek} F.\ Wilczek, {\it Two applications of axion
    electrodynamics}, {\sl Phys.\ Rev.\ Lett.} {\bf 58} (1987)
  1799--1802.

\bibitem{YakovGRG} Y.~Itin, {\it Coframe energy-momentum current.
    Algebraic properties,} Gen.\ Rel.\ Grav. {\bf 34} (2002)
  1819--1837; arXiv.org/gr-qc/0111087.

\bibitem{NP94} J.F.\ Nieves and P.B.\ Pal, {\it The third
    electromagnetic constant of an isotropic medium}, {Am.\ J.\ Phys.}
  {\bf 62} (1994) 207--216.

\bibitem{pureconn1} A.S.~Eddington, {\it The Mathematical Theory of
    Relativity}, 2nd ed.\ (Cambridge University Press, Cambridge,
  England, 1924).  

\bibitem{pureconn2} E.~Schr\"odinger, {\it Space-Time Structure}
  (Cambridge University Press, Cambridge, England, 1960), reprinted
  with corrections. 

\bibitem{pureconn3} D.~Catto, M.~Francaviglia, and J.~Kijowski, {\it A
    purely affine framework for unified field theories of
    gravitation,} {\sl Bull.\ Acad.\ Pol.\ Sci.\ (Phys.\ Astron.)}
  {\bf 28} (1980) 179--186.

\bibitem{pureconn4} F.~Gronwald, U.~Muench, A.~Macias, and F.W.~Hehl,
  {\it Volume elements of spacetime and a quartet of scalar fields},
  {\sl Phys.\ Rev.} {\bf D58} (1998) 084021 (4 pages); more complete
  in arXiv.org/gr-qc/9712063.

\bibitem{PRs} F.W.~Hehl, J.D.~McCrea, E.W.~Mielke, and Y.~Ne'eman:
  {\it Metric-Affine Gauge Theory of Gravity: Field Equations, Noether
    Identities, World Spinors, and Breaking of Dilation Invariance}.
  Phys. Rep. {\bf 258} (1995) 1--171.

\bibitem{Milutin} M.~Blagojevi\'c, {Gravitation and Gauge Symmetries}
  (IOP Publishing: Bristol, 2002).

\bibitem{gron96} F.~Gronwald and F.\,W.~Hehl, {\it On the gauge
    aspects of gravity}, in: {\sl Proc.\ Int.\ School of Cosm.\ \&
    Gravit.} 14\({}^{{\rm th}}\) Course: Quantum Gravity, held May
  1995 in Erice, Italy.  Proceedings. Erice, May 1995. P.G.  Bergmann
  et al.\ (eds.). World Scientific, Singapore (1996) pp.\ 148--198;
  arXiv.org/gr-qc/9602013.

\bibitem{Trautman} A.~Trautman, {\it The Einstein-Cartan theory}, in:
  {\sl Encyclopedia of Mathematical Physics}, J.P.~Fran\c{c}oise et
  al.\ (eds.). Elsevier, Oxford, 13 pages, to be published (2005), see

  {\tt http://www.fuw.edu.pl/}$\;\widetilde{\null}\;${\tt amt/ect.pdf} .

\bibitem{dil1} H.T.~Nieh, {\it A spontaneously broken conformal gauge
    theory of gravitation}, {\sl Phys.\ Lett.} {\bf A88} (1982)
  388--390.

\bibitem{dil2} T.~Dereli and R.W.~Tucker, {\it Weyl scaling and spinor
    matter interactions in scalar-tensor theories of gravitation},
  {\sl Phys.\ Lett.} {\bf B110} (1982) 206--210.

\bibitem{dil3}Yu.N. Obukhov, {\it Conformal invariance and space-time
    torsion}, {\sl Phys.\ Lett.} {\bf A90} (1982) 13--16.

\bibitem{IsmoIB} I.V.~Lindell, {\it The class of IB-media,} Helsinki
  Univ.\ Tech., Electromagnetics Lab.\ Report 459, 14 pages (June
  2005).

\bibitem{LindSihv2004a} I.V. Lindell, A.H. Sihvola, {\it Perfect
    electromagnetic conductor,} {\sl J.\ Electromag.\ Waves Appl.}
  {\bf 19} (2005) 861--869; arXiv.org/physics/0503232.

\end{thebibliography}
\end{document}